\documentclass[onecolumn,usenatbib,usegraphicx]{mn2e}
\def\FF{{\rm FF}}
\def\Mpc{{\rm Mpc}}
\def\lcdm{$\Lambda {\rm CDM} \,\,$} 
\def\mnras{MNRAS}
\def\aj{AJ}
\def\apj{ApJ}
\def\apjl{ApJL}
\def\aap{A\&A}
\def\pasj{Publications of the Astronomical Society of Japan}
\def\procspie{Proceedings of the SPIE}

\usepackage{epsfig}

\begin{document}
\title{ A 2D analysis of percolation and filamentarity in the  SDSS DR1}  
\author[B. Pandey and S. Bharadwaj ] { Biswajit Pandey\thanks{Email: 
    pandey@cts.iitkgp.ernet.in} and Somnath Bharadwaj\thanks{Email: 
    somnathb@iitkgp.ac.in} 
\\ Department of Physics and Meteorology\\ and
\\ Centre for Theoretical Studies \\ IIT Kharagpur \\ Pin: 721 302 ,
India }
\maketitle
\begin{abstract}
\end{abstract}
We have used the Largest Cluster Statistics and the Average
Filamentarity to quantify respectively the connectivity and the shapes
of the patterns seen in the galaxy distribution in two volume limited 
subsamples extracted from the  equatorial strips of the Sloan
Digital Sky Survey (SDSS) Data Release One (DR1). The data was
projected onto the equatorial plane and analyzed in two dimensions
(2D). Comparing the results with  Poisson point distributions at
various levels of smoothing  we find  evidence for a network like
topology with filaments being the dominant patterns in
the galaxy distribution. With increasing smoothing, a  transition from
many individual filamentary  structures to an interconnected network 
is found to occur at a filling factor $0.5-0.6$. We have tested the
possibility that the connectivity and the morphology of the patterns
in the galaxy distribution may be luminosity dependent   and   find
significant  evidence for a luminosity-morphology relation, the
brighter galaxies exhibiting lowers levels of connectivity and
filamentarity compared to the fainter ones.  Using a statistical
technique, Shuffle, we show that the filamentarity  in both the SDSS
strips  is statistically  significant up to $80 \, h^{-1} {\rm Mpc}$
but not beyond. Larger filaments, though identified, are not
statistically significant. Our findings reaffirm earlier work
establishing the filaments to be the largest known statistically
significant coherent structures in the universe. 


\section{Introduction}
Quantifying the clustering pattern observed in the galaxy distribution 
is one of the central themes  in modern cosmology. A striking 
feature visible in all   redshift  surveys   is that the
galaxies  appear to be distributed along filaments which are 
interconnected and form a network,  often referred to  as  the
``Cosmic Web''.  In this paper we quantify the
inter-connectivity and the shapes   of the patterns  seen in the
galaxy distribution in the SDSS (\citealt{york}).

The percolation analysis  (eg. \citealt{shand1};\citealt{einas1}) 
and the genus statistics (eg. \citealt{gott}) are some  of the earliest  
statistics introduced to   quantify the topology of the galaxy
distribution. \citet{shand2} used the Largest  
Cluster Statistics (LCS), a percolation  technique  developed for  
point-wise distributions,  to analyze the connectivity of structures in
the Las Campanas Redshift Survey (LCRS; \citealt{shect}). 
The thickness of the six LCRS slices is very small compared to the two
other  dimensions, and the analysis was carried out on two dimensional
(2D) projections. The $\rm LCS$ analysis focuses on the growth of the
largest cluster with increasing smoothing. A growth faster than that
of  a random Poisson point distribution indicates a network topology  
whereas a slower growth indicates a meatball topology. The  $\rm LCS$
analysis  shows  the presence of a  high level of inter-connectivity 
indicative of  a network like structure in the  distribution of the
LCRS galaxies. 

The Minkowski functionals have  been suggested   as a novel tool to
study the morphology  of structures in the universe (eg. \citealt{mecke}, 
\citealt{smal}). Ratios of the Minkowski functionals can be used to
define   a shape diagnostic 'Shapefinders' which faithfully quantifies
the  shapes of both simple and topologically complex objects
(\citealt{sahni}). \citet{bharad1}  defined the Shapefinder statistics
in 2D,  and  have used  this   to demonstrate that
the galaxy distribution in the LCRS exhibits a high degree of
filamentarity compared to a random Poisson  distribution. In a later
paper, 
\citet{bharad2} used Shapefinders 
in conjunction with a statistical technique  called Shuffle
(\citealt{bhav}) to determine the maximum length-scale at  which the
filaments observed in the LCRS are statistically significant.
They found that the largest length-scale at which filaments are
statistically significant is between  $70$ to $80 \,
h^{-1}$Mpc, for the LCRS $-3^o$ slice. Filamentary 
features longer than $80 \, h^{-1}$Mpc, though identified, are not
statistically significant. Such features arise from chance
alignments of galaxies. Further,  for the five other LCRS slices,
filaments of lengths $50 \, h^{-1}$Mpc to  $70 \, h^{-1}$Mpc were
found to be statistically significant, but not beyond.  

The ability to produce the  filamentary features observed in redshift
surveys  is an important test of any model for structure
formation. \citet{bharad3} have used N-body simulations
of the \lcdm  model with a featureless, scale invariant primordial
power spectrum and random initial phases to investigate  whether the  
filamentarity predicted by this model is consistent with that detected
in the LCRS. They find that the filamentarity in an  unbiased \lcdm
model is less than the LCRS. Introducing  a bias $b=1.15$,  the model
is in  rough consistency  with the data, and  a large bias  $(b=1.5)$
which enhances filamentarity at small scales and suppresses it at
large scales  is ruled out. The filamentarity is very
sensitive to the bias , and it may be possible to use a   quantitative
analysis of filamentarity to determine the bias parameter.

The  Sloan Digital Sky Survey (SDSS) offers the possibility of
studying coherent structures  in the galaxy distribution over
length-scales  which are substantially larger than  possible with 
earlier surveys like the LCRS.  Among the total sky coverage of the
publicly available 
SDSS Data Release One (DR1) are two strips which are centered along the
celestial equator ($\delta=0^{\circ}$), one  spanning  $65^{\circ}$
and the other $91^{\circ}$ in {\it r.a.},   and their thickness varying
within $\mid \delta \mid \le 2.5^{\circ}$ in {\it dec.}.  We use  galaxies
distributed over the redshift range $ 0.02  \le z \le 0.2$. The
thickness of the three dimensional regions corresponding to these 
strips is much smaller than the two other dimensions, and we can
project the galaxy distribution onto the equatorial plane $(\delta=0)$
without smearing out the large-scale patterns. In this paper we
quantify the inter-connectivity and the shapes of the patterns in the
resulting 2D  galaxy distribution using the same techniques which
were earlier applied to the LCRS.  

We restrict our analysis to volume limited
subsamples described in  Section 1 of this paper. The dense
sampling of the SDSS allows us to divide the galaxies into two classes
based on their luminosities and test if the inter-connectivity and the
shapes of the patterns seen in the galaxy distribution are different
for the fainter and the brighter galaxies. The large volume 
of the two SDSS equatorial strips offers further advantages over the
LCRS, we discuss these in the appropriate parts of the paper. In this
paper we have only considered the average properties of the
structures, and we have not addressed questions  pertaining to individual
filaments, nor have we compared our results to N-body simulations. It
is proposed to address these issues in future. 

\citet{hoyle1} have studied the 2D topology of the SDSS using the
genus statistics. In addition to the two equatorial strips used 
in this paper, they have also used a strip at high declinations. They
find that their results are consistent with those from  \lcdm
N-body simulations and also with those from a similar analysis of the
2dFGRS (\citealt{hoyle2}). They have also divided the
galaxies by colour and separately analyzed them to  find that the
distribution of the red galaxies shows a shift to a meatball topology
relative to the blue galaxies and the full sample, reflecting the fact
that red galaxies are distributed in more compact, high density
regions. \citet{hik} have  used Minkowski Functionals to
study the morphology of the patterns in the galaxy distribution in a
preliminary sample from the SDSS. \citet{doro} have used
the Minimal Spanning Tree to identify sheets and filaments in the SDSS
DR1. In both these works the authors find their  results to be
consistent with \lcdm N-body simulations.\citet{einas2}
have studied the super-cluster void network in the SDSS. \citet{sheth2}
has used a technique SURFGEN (\citealt{sheth1}) to study  the
geometry, topology and morphology of the superclusters in mock SDSS
catalogues and find  that the filamentarity is the dominant
morphology of the large superclusters. \citet{shand3} studied the
large-scale network in the dark matter density field in N-body
simulations from VIRGO consortium using SURFGEN and noticed that the
individual superclusters and voids exhibits a significant amount of
substructures as indicated by their genus values.  

\citet{basil} and \citet{koloko} have studied the super-cluster void
network in the PSCz and the Abell/ACO cluster catalogue respectively, finding
filamentarity   to be the dominant feature. \citet{pimb} performed an
analysis of the frequency and distribution of intercluster filaments
in the 2dfGRS with a filament classification scheme based on their
visual morphology and reported that massive clusters have larger
number of filaments.

Traditionally, correlation functions  have been used to quantify the
statistical properties of the galaxy distribution with  the  two-point 
correlation function   and its Fourier transform, the power
spectrum, receiving most of the attention.   For the SDSS this
includes analysis of the power spectrum(eg. \citealt{szal};
\citealt{teg1}; \citealt{teg2}; \citealt{dodel}), the two-point
correlation function  (eg. \citealt{zevi}; \citealt{conn};
\citealt{infan} ) and the higher order moments (eg. \citealt{szap}).  

We have used a \lcdm cosmological model with $\Omega_{m0}=0.3$,
$\Omega_{\Lambda0}=0.7$ and $h=1$ throughout. 

We next present an outline of the paper. In Section 2 we describe the
data and the method of analysis. The results are presented in Section
3, and finally we present discussion and conclusions in Section 4.

\section{Data and method of analysis}
\subsection{SDSS and the data}
The SDSS  is a wide-field photometric  and spectroscopic survey of the 
high galactic latitude sky visible from  the Northern hemisphere. It
uses a dedicated 2.5 m telescope at Apache 
Point Observatory in New Mexico. The primary goals of the SDSS are to
image 10,000 square degrees of the Northern Galactic Cap and three
$\sim 200$ square degree stripes in the Southern Galactic Cap  
in five wavebands namely u, g, r, i and z, and determine   spectroscopic
redshifts of approximately $10^8$ galaxies and $10^5$ 
quasars.  
A high level overview of the SDSS is provided by \citet{york}. The
details of the software and data products of the Early Data
Release(EDR) are described in \citet{stout}. The details and
the updates of the data for the  Data Release One (DR1) and Second
Data Release (DR2) can be found in two papers,  \citet{abaz1}
and \citet{abaz2} respectively. Other technical details of the
SDSS are the descriptions of the photometric camera (\citealt{gunn}), the
photometric system (\citealt{fuku}; \citealt{smith}), photometric
monitor (\citealt{hog}) and photometric analysis (\citealt{lup}). 
There are other important articles which covers astrometric
calibrations (\citealt{pier}), selection of spectroscopic
samples (\citealt{eisen}; \citealt{stras}) and spectroscopic
tiling (\citealt{blan}). 

Our present analysis is based on SDSS DR1 galaxy redshift data. 
In this paper we analyze two equatorial strips (celestial
equator) one in Northern Galactic  Cap (NGP) which covers the region
$145^\circ<\alpha<236^\circ$, and another in the Southern Galactic Cap
(SGP) covering $351^\circ<\alpha<56^\circ$, both with varying
thickness in the range $-2.5^\circ<\delta<2.5^\circ$.  This contains
38,838 galaxies having redshift in the range $0.02\leq z \leq 0.2$
with the selection criteria that  the extinction corrected Petrosian r
band magnitude is $r_p < 17.77$.   

\begin{figure}
\rotatebox{-90}{\scalebox{.6}{\includegraphics{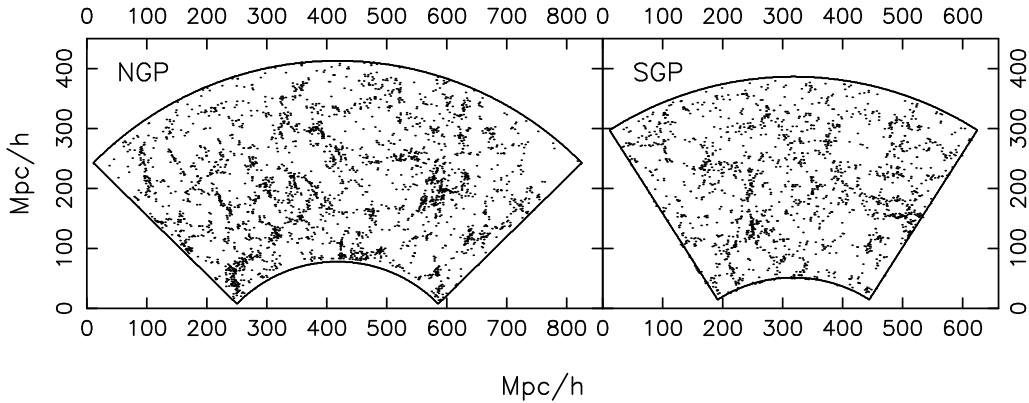}}}
\caption{This shows the galaxy distribution  in two of the volume
 limited subsamples which we have analyzed.  The other subsamples, 
  described in Section 2.1 and listed in
 Table I, were all extracted   from the two subsamples shown here
 and  the visual appearance of these subsamples is very similar.  
}
\label{fig:1}
\end{figure}

\begin{figure}
\rotatebox{-90}{\scalebox{.6}{\includegraphics{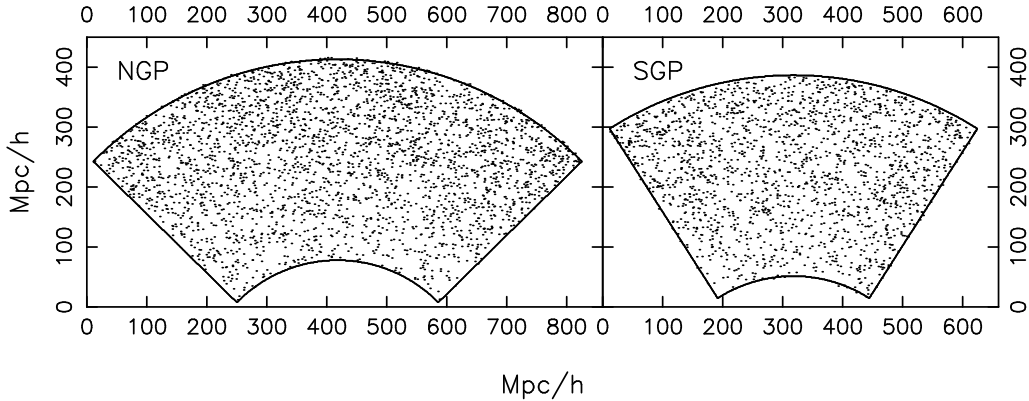}}}
\caption{This shows a single realization of the  random galaxy
  distribution  in  two of the  SDSS strips  that we have analyzed. 
}
\label{fig:2}
\end{figure}

For the current purpose we  selected only the galaxies lying within
 $-1^\circ<\delta<1^\circ$ as both the equatorial strips have complete
 coverage in this declination range.   
We construct a volume limited samples over the redshift range
$0.08\leq z \leq 0.2$  by restricting the extinction corrected
Petrosian r band apparent magnitude in the range $14.5\leq m_r \leq
 17.5$ and absolute  magnitude in the range $-22.6\leq M_r \leq
 -21.6$.     
This reduces  the number of galaxies, but also offers some advantages
as the radial selection function is approximately uniform  so the
variation in number density of galaxies is caused  by clustering only.  
The above redshift limit was chosen so as to get a good compromise
 between the number of galaxies and the volume of the sample.We
 further divide the absolute magnitude range into two equal parts in 
order to produce separate
volume limited samples of the  fainter and brighter galaxies. We
finally have 5315 galaxies distributed in two wedges, spanning
$91^{\circ}$ (NGP) and $65^{\circ}$ (SGP) in ${\it r.a.}$, both with
thickness $2^{\circ}$ centered along the equatorial plane extending
 from $235 \, h^{-1} \Mpc$ to $571 \, h^{-1} \Mpc$ comoving in the
 radial  direction. 

 The analysis using Shuffle  requires us to cut the entire  survey
 area into squares and shuffle them around.  The thickness of the
 wedges described above increases with radial distance varying from 
$8.2 \, h^{-1} \, \Mpc$ to $20 \, h^{-1} \Mpc$. For the
 Shuffle analysis we have used   subsamples of uniform thickness $8.2 \,
 h^{-1} \, \Mpc$ extracted from the wedges described above. 

 Table I summarizes some properties of all the subsamples
 which we have used in our analysis. The galaxy positions in all the
 subsamples  were 
 projected onto the equatorial plane to obtain  the 2D galaxy
 distributions  (Figure \ref{fig:1}) which  we have analyzed in 
 the rest of the  paper. A visual inspection of Figure \ref{fig:1}
 reveals the presence of large-scale coherent patterns namely the
 interconnected network of filaments and voids which we now proceed to 
 quantify. For both the NGP and SGP subsamples (Table I), we have
 generated 9 random realizations (Figure \ref{fig:2}) which contain
 exactly the same number of galaxies randomly distributed over the
 same volume  as the actual data. 

\begin{center}
\begin{table}{Table I}\\
\begin{tabular}{|c|c|c|}
\hline
Sample & Abs. Mag &   No. of galaxies \\
\hline
NGP & $-22.6\leq M_r \leq -21.6$ & $3297$ \\
NGP faint & $-22.1\leq M_r \leq -21.6$ &  $2221$ \\ 
NGP bright & $-22.6\leq M_r < -22.1 $ & $1076$ \\ 
NGP uniform thickness & $-22.6\leq M_r \leq -21.6$ & $1936$ \\
SGP & $-22.6\leq M_r \leq -21.6$ & $2018$ \\
SGP faint & $-22.1\leq M_r \leq -21.6$ &  $1206$\\ 
SGP bright & $-22.6\leq M_r < -22.1$ & $812$\\ 
SGP uniform thickness & $-22.6\leq M_r \leq -21.6$ & $1096$ \\
\hline
\end{tabular}
\end{table}
\end{center}

\subsection{Method of Analysis}

For each of the subsamples described in the previous subsection,   
the 2D galaxy distribution was
 embedded in a $1 \, h^{-1} {\rm Mpc} \times 1 \,h^{-1} {\rm  Mpc}$
 2D rectangular  grid. Grid cells having a galaxy within them are
termed as filled  and were assigned the value 1,  whereas cells with
 no  galaxy are termed as empty and were given the  value 0. 
The grid cells which are beyond the boundaries  of the survey were
 assigned a negative value in order to distinguish them from the empty
 cells within the  survey area. The net result of is that  the galaxy
 distribution is now represented through a  distribution of  1s and 0s
 located on a 2D grid.   

The next step is to use an objective criteria to  identify the
coherent large-scale structures visible in the galaxy distribution.    
We use a ``friends-of-friends''(FOF) algorithm to identify interconnected  
regions of  filled cells which we refer to as clusters. In this
algorithm any two adjacent filled cells are referred to as friends. 
Clusters are defined through the stipulation that any friend of my
friend is my friend. The distribution of 1s on the grid is very sparse
with only $\sim 1 \%$ of the cells being  filled.  Also, the filled
cells  are mostly isolated, and the clusters identified using  FOF,
which contain only a few cells each, do not correspond to the
large-scale  coherent structures  seen in the SDSS strips. It  is
necessary to  smoothen or coarse-grain the galaxy distribution so 
that the large scale structures may be objectively identified. 

\begin{figure}
\rotatebox{0}{\scalebox{.8}{\includegraphics{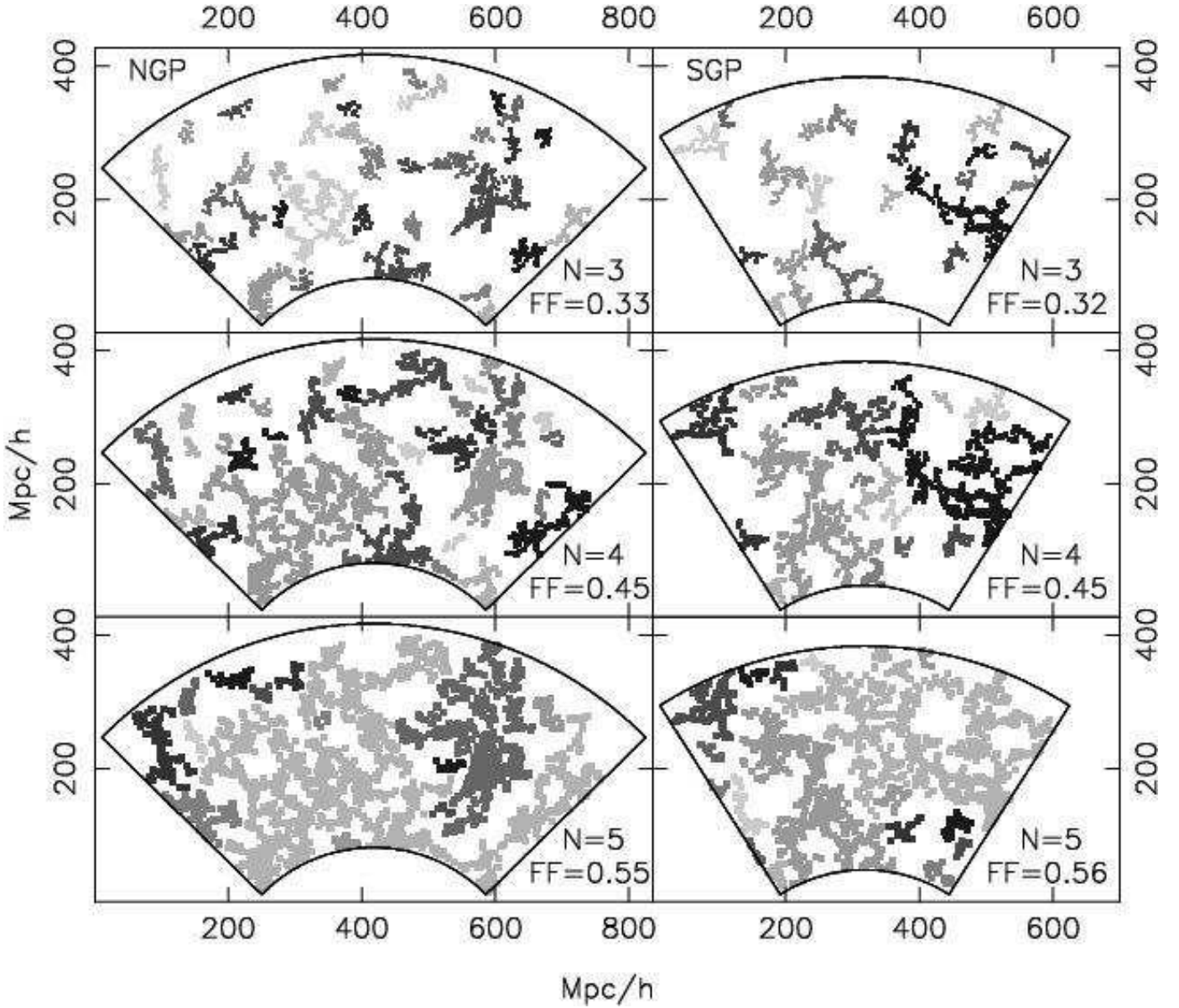}}}
\caption{This shows the NGP and SGP strips at different levels of
  coarse-graining indicated by $N$ in the figure along with the
  corresponding filling factor $FF$. Only the large clusters are
  shown, different shades (colours) being used to demarcate individual
  clusters.  
}
\label{fig:3}
\end{figure}

The coarse-graining is carried out by gradually  making each filled
cell fatter until the filled cells overlap and they finally  fill up the
whole survey area. 
In every iteration of coarse-graining we fill up all the empty cells
which are adjacent to filled cells, causing  every  filled cell to  grow
fatter. This causes clusters to grow, first because of the growth of
filled cells, and then by the merger of adjacent clusters as they
overlap. This process is illustrated in Figure \ref{fig:3}.  
At the initial stages of 
coarse-graining,  the patterns which emerge from the distribution of
1s and 0s closely resembles the coherent  structures  seen in the
galaxy distribution. As the coarse-graining proceeds, the clusters become
very thick and fill up the entire region washing away any visibly
distinct pattern. The filling factor $\FF$, defined as the fraction of
cells within the survey area that are filled, {\it ie.}
\begin{equation}
\FF=\frac{{\rm Total \, No.\,  of\,  Filled \, Cells}}{{\rm Total \,
 No.\,  of \, Cells \, Inside \, the \,  
 Survey \, Area}}
\end{equation}
 increases from $\FF \sim$ 0.01 to $\FF =$ 1 as the coarse graining
 proceeds. So as to not 
restrict our analysis to an arbitrarily chosen level of smoothing, we
analyze the clusters identified in the pattern of 1s and 0s after each
iteration of coarse-graining {\it ie.} the whole range of $\FF$. 

\subsubsection{LCS and Shapefinders}

At each level of coarse-graining we identify the largest cluster and
calculate  the Largest Cluster Statistic (LCS) defined as the fraction of
the filled cells in the largest cluster 
\begin{equation}
{\rm LCS}=\frac{{\rm No.\,  of \, Filled \, Cells \, in \, the \, Largest
    \, Cluster}}{{\rm Total \, No. \, of \, Filled \, Cells}} \,.
\end{equation}
We study the growth of LCS with increasing FF (Figures \ref{fig:5} and
\ref{fig:6}) to quantify the tendency of clusters to get
interconnected as the coarse-graining proceeds (Figure
\ref{fig:3}). The transition from many small clusters to a 
network of interconnected filaments running across almost the entire
survey, which is the onset of percolation, is manifested by a sharp
increase in LCS.

The geometry and topology of
a two dimensional cluster can be described by the three Minkowski
functionals, namely its area $S$, perimeter $P$, and genus $G$. It is
possible to quantify the shape of the cluster using a single  2D
``Shapefinder'' statistic  (Bharadwaj et al. 2000) which is defined 
as the dimensionless ratio
\begin{equation}
{\cal F}=\frac{P^2 - 4 \pi S}{P^2 + 4 \pi S}\,, 
\end{equation}
which by construction  has values in the range $0 \le {\cal F} \le 1$. It
  can be verified that  ${\cal 
  F} =1$ for an ideal filament which has a finite length and zero
  width, whereby it subtends no area ($S=0$) but has a finite
  perimeter ($P>0$). It can be further checked that ${\cal F}=0$ for a
  circular disk, and intermediate values of ${\cal F}$ quantifies the
  degree of filamentarity with the value increasing as a cluster is
  deformed from a circular disk to a thin filament. 

The definition of ${\cal F}$ needs to be modified when working on a
rectangular  grid of spacing $l$.  An ideal filament, represented on a
grid,   has the minimum possible width {\it i.e.} $l$,  and its perimeter
$P$ and area $S$ are related as 
$P=2 S + 2 l$. At the other extreme we have  $P^2=16 S$ for a square 
shaped cluster on the grid. We introduce the  2D Shapefinder statistic 

\begin{equation}
{\cal F} = \frac{(P^2 - 16 S)}{(P-4 l)^2}
\end{equation}
to quantify the shape of  clusters on a grid.  By definition 0$\le
{\cal F} \le$ 1.  ${\cal F}$ quantifies the degree of filamentarity of the 
cluster, with ${\cal F}$ = 1 indicating a filament  and ${\cal F}$ =
0, a square, and ${\cal F}$ changes from $0$ to $1$ as a square is
deformed to a filament.

The extent of filamentarity in a survey is mostly dominated by the
morphology of the most massive members as a large cluster has a greater
contribution to the overall texture of large-scale structure than an
individual galaxy.We therefore want to give greater weight to larger
objects and used the second moment of filamentarity as the indicator
of average filamentarity.
The average filamentarity $F_2$ is defined as the mean filamentarity
of all the clusters in a slice weighted by the square of the area of
each  clusters

\begin{equation} 
F_2 = {\sum_{i} {\cal S}_i^2 {\cal F}_i\over\sum_{i}{\cal S}_i^2} \,. 
\end{equation}

In the current analysis, we study  the average filamentarity $F_2$ as
a function of FF (Figures \ref{fig:5} and \ref{fig:6}) to quantify the
degree of filamentarity in each of  the SDSS subsamples and the random
datasets. 

\subsubsection{Shuffle}
\begin{figure}
\rotatebox{0}{\scalebox{.6}{\includegraphics{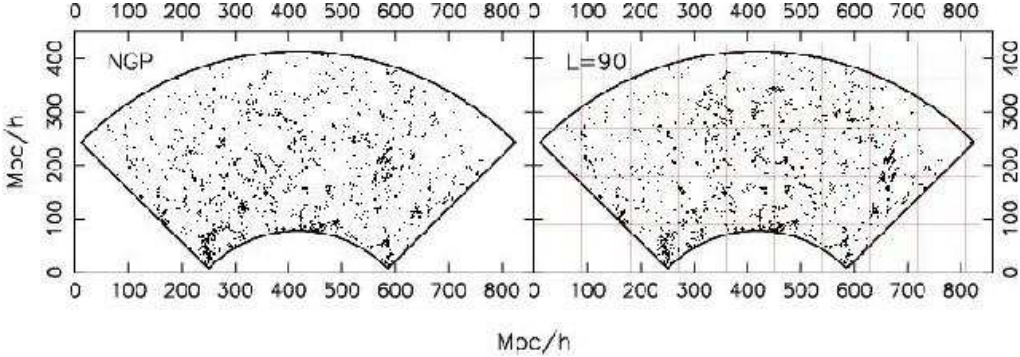}}}
\caption{ This figure exhibits how Shuffle works.  The left panel
  shows the galaxy distribution in the NGP  
  subsample and the right panel shows a  shuffled realization generated
  from  the same data using   $L=90 h^{-1} {\rm Mpc}$. A $90
  h^{-1}{\rm Mpc} \, \times \, 90 h^{-1} {\rm Mpc}$ grid was placed on
  the NGP data, and  square  blocks lying fully inside the survey
  region   were randomly shuffled   around to generate the data shown
  in the   right panel.  This process destroys all coherent structures
  spanning   length-scales larger than $L=90 h^{-1} {\rm Mpc}$ in the
  actual   data and filaments larger than this in the shuffled data
  arise from   chance alignments. Among the filaments seen in the
  right panel,   those which  run across the block  boundaries have
  formed purely   from chance alignments. 
}
\label{fig:4}
\end{figure}

We use a statistical technique called Shuffle to determine the largest
length-scale at which the filamentarity is statistically significant. 
A grid with squares blocks of side $L$ is superposed on the original
data 
slice (Figure \ref{fig:4}). Blocks of data which lie entirely
within the slice  are then randomly  interchanged, with rotation,
repeatedly, to form a new shuffled  slice. The shuffling process
eliminates coherent features in the original data on scales larger than
$L$, keeping clustering at scales below $L$ nearly identical to  the
original data.   All the structures spanning length-scales greater
than $L$ that exist in the shuffled slices are the result of chance
alignments. At a fixed value of $L$, the average filamentarity
in the original sample will be larger than  in the shuffled data only
if the actual data has more filaments spanning length-scales larger
than $L$, than that expected from chance alignments. The largest value
of $L$, $L_{\rm MAX}$, for which the average filamentarity of the
shuffled slices is less than the average filamentarity of the actual
data gives us the largest length-scale at which the filamentarity is
statistically significant. Filaments spanning length-scales larger
than $L_{\rm MAX}$ arise purely from chance alignments.

We have used the uniform thickness subsamples (Table I) which have
thickness $8.2 \, h^{-1} \, {\rm Mpc}$ for the Shuffle analysis. For
each value of $L$ we generated $24$ different  realization of the
shuffled slices.  To ensure that the edges of the  blocks which are
shuffled around do  not cut the actual filamentary pattern at exactly
the same place in all the  realizations of the shuffled data, we
randomly shifted the origin of the grid used to define the blocks.  
The values of FF and $F_2$ in the 24 realizations differ from one
another and from the actual data at the same stage of
coarse-graining. So as to be able to quantitatively compare the
shuffled realizations with the actual data, we interpolate the values
of $F_2$ in the shuffled realization at the values of FF obtained for
the actual data.  The mean $\bar{F_2}[{\rm Shuffled}]$ and the
variance  $(\Delta F_2[{\rm Shuffled}])^2$ of the average
filamentarity was determined for the shuffled data at each value of FF
using  the 24 realizations.  The  difference between the filamentarity
of the 
shuffled data and the actual data was quantified using  the reduced
$\chi^2$ per degree of freedom  
\begin{equation}
\frac{\chi^2}{\nu}=\frac{1}{N_p} \sum_{a=1}^{N_p}
 \frac{(F_2[{\rm Actual}]-\bar{F_2}[{\rm Shuffled}]))_a^2 }
{(\Delta  F_2[{\rm Shuffled}])_a^2} 
\end{equation}
where the sum is over different values of the filling factor FF. 

\section{Results}
\subsection{Comparison with random samples}
\begin{figure}
\rotatebox{-90}{\scalebox{.6}{\includegraphics{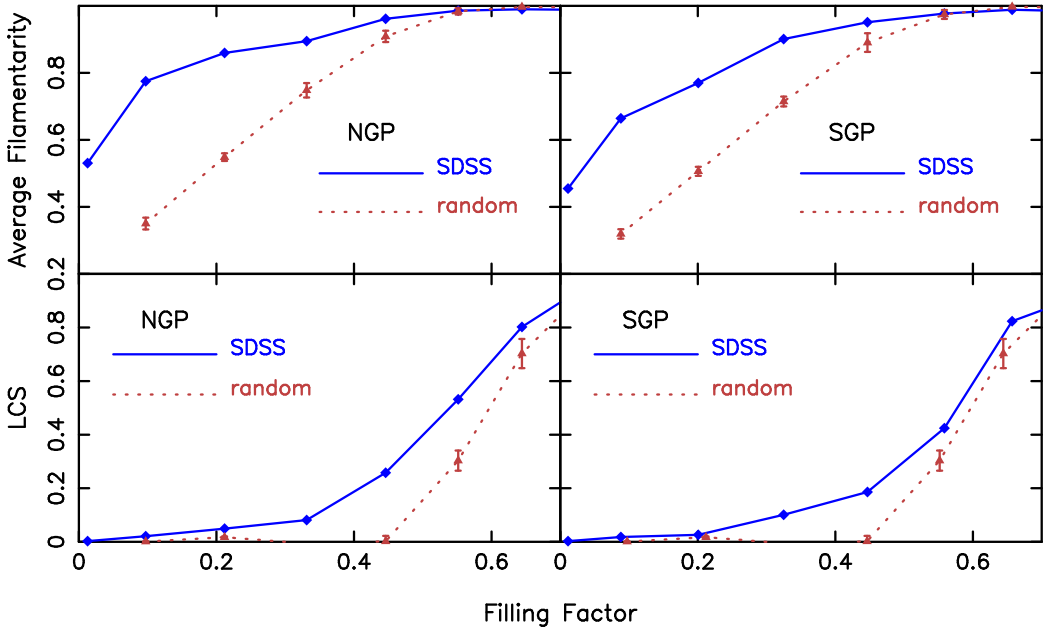}}}
\caption{This shows the Largest Cluster   Statistics (LCS) and the
  Average Filamentarity ($F_2$)  for two of the SDSS strips together
  with the values for their random counterparts. We have used 9
  realizations  to determine the mean values and
  the $1-\sigma$ error-bars shown  for the  random data.  
}
\label{fig:5}
\end{figure}

Figure \ref{fig:5} shows the results for the NGP and SGP strips
plotted along with the values for their random counterparts.  For both
NGP and SGP we have  9 random realizations which were analyzed in
exactly the same way as the actual data. The values of the filling
factor FF  differ from realization to realization at the same level of
coarse-graining, and we interpolated the values of LCS and  $F_2$ at
the values of FF obtained for the actual data, and these were used to
obtain the mean and the $1-\sigma$ error-bars shown for the random
data in the figure.  It may be noted that the values of LCS and $F_2$
show very little variation from  realization to realization,  and this
is reflected in the very small error-bars. This is a consequence of
the  high number density and large area of the two SDSS strips
analyzed here. 

For both the actual data and the random realizations,  the LCS, has a
very small value at low values of FF. For the actual data, LCS is
below $0.2$ up to a filling factor of $0.4$ {\it ie.} the largest
cluster contains less than $20 \%$ of the filled cells when around $40
\%$ of the cells in the survey area are filled. A 
transition is seen to occur at FF in the range $0.5-0.6$ for  both
the NGP and SGP strips,  and the largest cluster contains more than $
80 \%$ of all the filled cells at a filling factor  ${\rm FF}=0.65$. This
sharp transition from a set of small clusters at ${\rm FF} < 0.4$ to a 
single large cluster which is a network of filaments running  across
almost the entire survey  is referred to as the percolation transition 
and  it is found to occur at a threshold value of the filling factor 
 around ${\rm FF}=0.5$.  The LCS of the random data is less than that
 of  the actual data  for nearly the entire range of FF, the two
 curves  being many factors of $\sigma$ apart.  Also, in the random
 data ${\rm LCS } <0.2$ all the way  to ${\rm FF}=0.5$ whence it
 exhibits a sudden rise to ${\rm LCS} =  0.7$ at ${\rm FF}=0.65$.   
  The faster growth of LCS with increasing FF in the two SDSS strips
  as compared to the random Poisson point distribution, and the onset
  of percolation at a lower value of FF in the actual data show the
  presence of network like topology in the galaxy distribution.  

Turning our attention next to the average filamentarity $F_2$, we find
that initially   $F_2$ is larger in the actual data as compared to the
random data.  For both the actual  and the random data $F_2$ increases
with successive iterations of coarse-graining and reaches a value $F_2
\sim 1$ at ${\rm FF}=0.6$. This corresponds to a situation where
nearly all the clusters have merged into a single cluster (${\rm LCS}
\sim 0.8$) and further smoothing results in only fattening this
cluster causing the average filamentarity to fall. The region beyond
${\rm FF}=0.6$ is not of importance and is not considered in our
analysis. The average filamentarity of the actual data is larger than
that for a  Poisson point distribution for the entire range of filling
factor ${\rm FF} \le 0.6$. This shows that the two SDSS strips analyzed
here are largely dominated by filaments,  significantly in excess of
that expected in a random distribution of points.

\subsection{Luminosity Dependence}
\begin{figure}
\rotatebox{-90}{\scalebox{.6}{\includegraphics{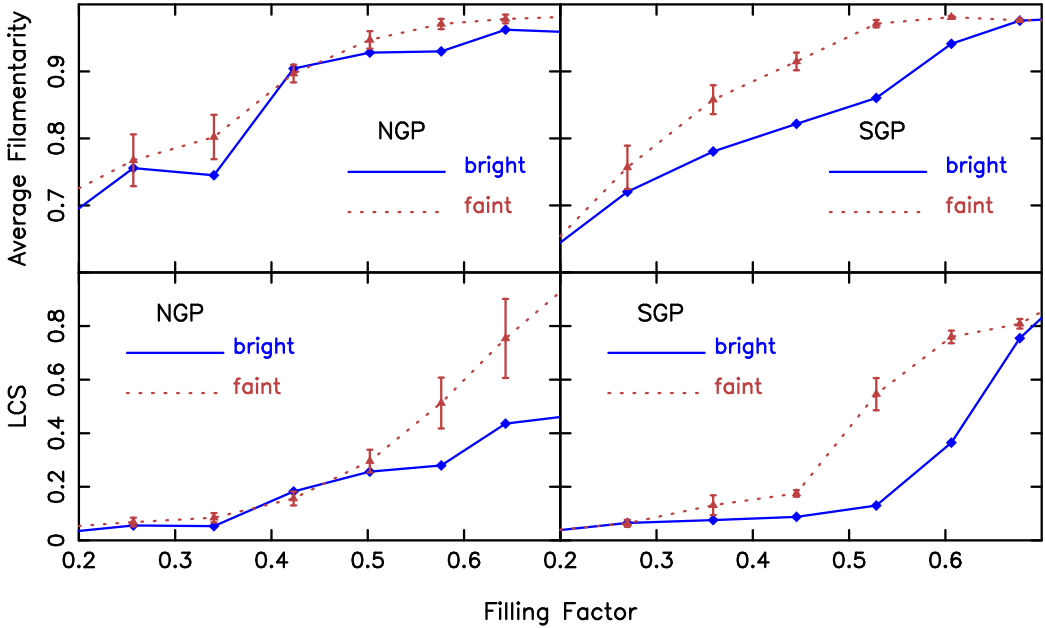}}}
\caption{The average filamentarity and largest cluster statistics
  vs. filling factor for
  bright and faint galaxies in the volume limited samples.The
  $1-\sigma$ error-bars for faint galaxies are shown in the figure.
}
\label{fig:6}
\end{figure}

The bright and faint subsamples were separately analyzed to
investigate the luminosity dependence, if any,  of the  connectivity
and the shapes of the patterns in the galaxy distribution. For both
the SGP and the NGP the subsample of faint galaxies (Table I) contains
roughly $1.5$ to $2$ times the number of galaxies in  the subsample of
bright galaxies distributed over the same region. So as to compare the
bright and faint subsamples at the same galaxy number density we 
randomly extracted a subset of the faint subsamples  so that the
number exactly matches the bright subsamples.  Five such randomly
chosen subsets were used  to make
bootstrap estimates of the mean and the $1-\sigma$ fluctuations of LCS
and $F_2$ as function of ${\rm FF}$ (Figure \ref{fig:6}).  These were
compared with LCS and $F_2$ for the bright subsample to test if there
is any statistically significant evidence for a  luminosity dependence. 

We find that the values of both LCS and $F_2$ are larger for the
faint subsamples as compared to the bright ones.  The percolation
transition occurs at a smaller value of $FF$ for the faint subsamples,
indicating that a network like topology is  more dominant in the
distribution of faint galaxies as compared to the 
bright ones. Also, the faint galaxies have a more  filamentary
distribution,  quantified by $F_2$,  as compared to the  bright 
galaxies. Using the reduced $\chi^2$ per degree of freedom to asses
the statistical significance of these differences we find that
$\chi^2/\nu=(72, 39)$ in  NGP and SGP respectively for the
Largest Cluster Statistics and   $(16,97)$  for the  Average
Filamentarity.  This establishes that the luminosity dependence is a
statistically significant effect.

\begin{figure}
\rotatebox{-90}{\scalebox{.6}{\includegraphics{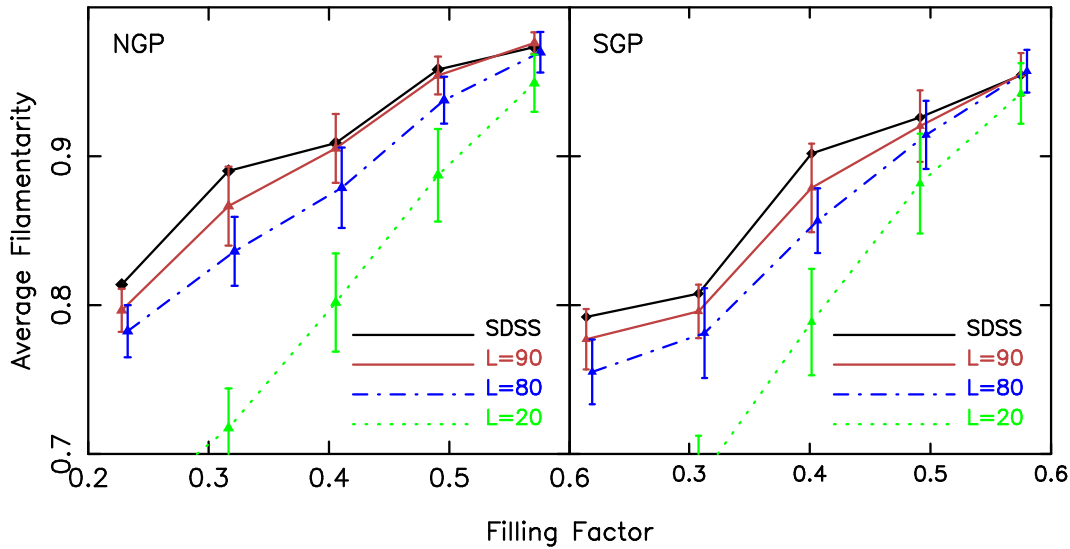}}}
\caption{This shows the  Average Filamentarity  for   the  two SDSS
  slices together with the results  for   the shuffled data for three
  values of $L$ shown in the figure.  Shuffling with $L=20 h^{-1} \,
  {\rm Mpc}$ causes a large drop in the Average Filamentarity showing
  the statistical significance of the filamentarity at this
  length-scale. The filamentarity is statistically significant up to
  $L_{MAX}= 80 h^{-1} \,  {\rm Mpc}$ where the actual data lies above
  the the $1-\sigma$ error-bars. The data is within the  $1-\sigma$ 
  error-bars of the shuffled   realizations for larger values of $L$,
  indicating that the filaments are not statistically   significant
  beyond $L_{MAX}$. The data point for $L= 80 h^{-1} \,  {\rm Mpc}$
  have been slightly shifted to prevent the error-bars from
  overlapping in the figure. 
 }
\label{fig:7}
\end{figure}
  
\begin{center}
\begin{table}{Table II}\\
\begin{tabular}{|c|c|c|}
\hline
L& $\chi^2/\nu$(NGP) & $\chi^2/\nu$(SGP)\\
\hline
20 & 16.64 &  15.1\\
30 & 11.7 &  5.2 \\
40 & 5.8&  4.53\\
50 & 5.3&  1.64\\
60 & 3.43&  2.85\\
70 & 2.2&  1.8\\
80 & 2.33&  1.65\\
90 & 1.08&  0.82\\
100 &  0.58 & 0.46\\
110 &  1.03& 0.66\\ 
120 & 1.13& 0.33\\
130 & 0.5& 0.5\\
140 & 0.5& 0.4\\
150 & 0.4& 0.3\\
\hline
\end{tabular}
\end{table}
\end{center}

\begin{figure}
\rotatebox{-90}{\scalebox{.6}{\includegraphics{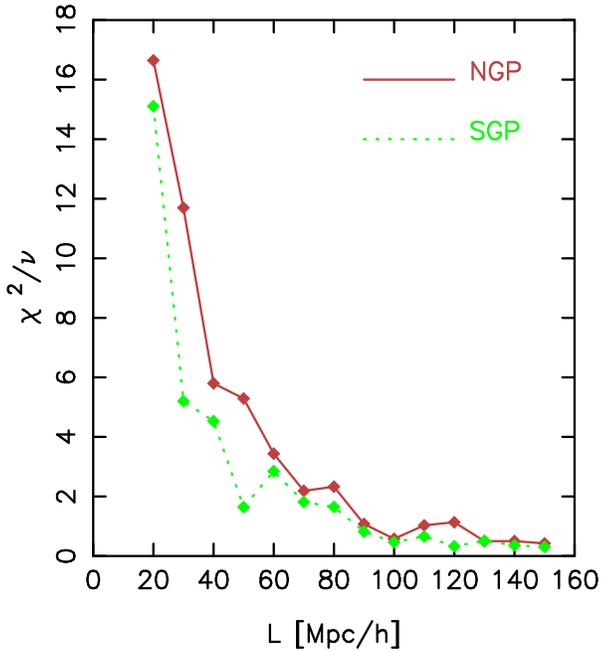}}}
\caption{This shows the $\chi^2/\nu$ at different 
  shuffling lengths $L$ for both the NGP and SGP slices. 
}
\label{fig:8}
\end{figure}

\subsection{Statistical Significance of the Filaments}
We applied Shuffle to the NGP and the SGP uniform thickness subsamples 
varying $L$ from $20 $ to $150 \, h^{-1} \, {\rm Mpc}$ in steps of $10
\, h^{-1} \, {\rm Mpc}$. The Average 
Filamentarity falls substantially if the data is Shuffled using $L\, = 
\,  20 \, h^{-1} \, {\rm Mpc}$ (Figure \ref{fig:7}) indicating that a
large fraction of the filaments are cut by the Shuffling mechanism and
the number of filaments which are produced by chance alignments in the
Shuffled data are less than the number of filaments destroyed. This
establishes that  the  filaments  do not arise from chance alignments
and are  statistically significant, genuine features of 
the   galaxy distribution  at this length-scale. 

Longer filaments survive the Shuffling process as $L$ is increased and
hence the  Average Filamentarity increases with $L$ , slowly
approaching the values for the actual data.  The values of
$\chi^2/\nu$ given in Table II and shown graphically in Figure
\ref{fig:8}  quantify the difference in $F_2$ between the Shuffled and
the actual data.  We  find that Shuffling the data with $L = 90 \,
h^{-1} \, {\rm Mpc}$ or larger doesn't result in a
statistically significant drop in $F_2$, the value of $\chi^2/\nu$
being $\sim 1$ for $L \ge 90 \, h^{-1} \, {\rm Mpc}$. The value of 
$L_{MAX}$, the largest length-scale at which Shuffling causes $F_2$ to
fall, is  $80 \, h^{-1} \, {\rm Mpc}$ for both NGP and SGP. This is
the largest length-scale at which the filaments are statistically
significant. 

Cutting the galaxy distribution into blocks of  size $90 \, h^{-1} \,
{\rm Mpc}$ or larger and shuffling them around  does not reduce the
Average Filamentarity. Filaments spanning  length-scales larger than 
$90 \, h^{-1} {\rm Mpc}$ are present in equal abundance in the Shuffled
  and the actual  data showing that these filaments  arise purely from
  chance alignments.  

The Average Filamentarity for $L=80 \, h^{-1} {\rm Mpc}$ and $90 \,
h^{-1} {\rm Mpc}$ where we have the transition from statistically
significant filaments to filaments that arise purely from chance is
shown in Figure \ref{fig:7} along with the actual data.  It may be
noted that we have restricted our analysis to $0.2 \le {\rm FF} \le
0.6$  as there are many small clusters  which bear no resemblance to
the filaments we wish to characterize for ${\rm FF}<0.2$ and nearly
all the filaments get interconnected into a single dominant cluster
for ${\rm FF}>0.6$.

\section{Discussion and Conclusion}
The large-scale network of filaments and voids is one of the
 most striking visual features in all galaxy redshift
 surveys e.g., CfA(\citealt{gel}), LCRS \citep{shect}, 
 2dFGRS(\citealt{colles}) and
 SDSS(EDR) (\citealt{stout}, \citealt{abaz1}). The SDSS, the largest
 redshift survey to date, offers a unique opportunity to study these
 features at a high level of precisions on hitherto unprecedented
 length-scales.    We have used the Largest  Cluster Statistics (LCS)
 and the Average Filamentarity ($F_2$) to  study respectively the
 interconnectivity  and the shapes of the patterns seen in the galaxy
 distribution in 2D  projections of volume limited subsamples from the
 two equatorial strips in the  SDSS DR1.  These were  studied as
 functions of the Filling Factor 
 (FF) at different levels of smoothing, and the results compared to a
 2D Poisson point distribution. We find that at the same value of FF,
 the values of LCS and  $F_2$ in both the NGP and the SGP strips are
 substantially   larger than those of the random distributions for
 nearly the entire range of FF. This indicates a high level of
 connectivity consistent with a network like topology with filaments
 being the dominant structures in the galaxy distribution. Individual
 filamentary structures  identified at low filling factors $({\rm FF}
 \sim 0.3)$  get interconnected into a network at the percolation
 transition $({\rm FF} \sim 0.5 - 0.6)$. 

The high number density of galaxies in the SDSS allows us to test if
there is evidence for luminosity dependence in the connectivity and
the 
filamentarity. We find that the distribution of the brighter galaxies
exhibits lesser  connectivity and filamentarity compared to the
fainter galaxies in the same region. This is consistent with the
picture where the brighter galaxies preferentially reside in the compact
high density regions whereas the fainter galaxies have a more diffuse 
distribution. Studies using N-body simulations \citep{bharad3}   
have shown that the filamentarity is highly sensitive to bias, with
the large-scale filamentarity falling with increasing bias.  The
findings of this paper  reaffirm that the brighter galaxies have a
higher bias relative to the fainter ones as noted from earlier studies
of the luminosity dependence of the galaxy clustering
(eg. \citealt{nor}, 
\citealt{zevi}). \citet{einas2} have studied the luminosity
distribution of galaxies in high and low density regions of the SDSS
to show that the brighter galaxies are preferentially distributed in
high density environments. \citet{goto} have studied the
morphology-density relation in the SDSS(EDR) and find that this
relation is less noticeable in the sparsest regions indicating
requirment of denser environment for the physical mechanisms
responsible for galaxy morphological change. 
Studies of the connectivity using the genus
statistics \citep{hoyle1} have revealed  colour dependence with the
redder galaxies showing a lesser connectivity  compared to 
the blue ones. \citet{hog1} and \citet{blan1} find  a strong 
environment  dependence for both the colour and  luminosity for the
SDSS galaxies. Their results indicate that the red galaxies are found
preferentially in overdense regions relative to the blue galaxies. 
These observations showing the connectivity and filamentarity to
depend on the luminosity and the color of the galaxies  poses interesting
questions about both, the models of galaxy formation and  our
understanding of the formation of the filament-void network. 
It may be noted that the colour dependence of the filamentarity has
not been studied here  and it is proposed to take this up in the
future.  

Studies of the genus statistics at large values of smoothing shows the
SDSS to be consistent with a Gaussian random field \citet{hoyle1}. Our
analysis of the connectivity starting  with the unsmoothed data and
analyzing it at various levels of smoothing complements the earlier
analysis and reveals the presence of strongly non-Gaussian features,
namely the filaments.  It is interesting to note that these filaments
are the natural outcome of gravitational instability starting from
Gaussian random initial conditions. 

  We have determined the largest length-scale at which the
  filaments are statistically significant. We find that filaments
  spanning length-scales up to $80 \, h^{-1} \, {\rm Mpc}$ are
  statistically significant in both the  SDSS strips we have
  analyzed. Filaments spanning scales larger than this are the outcome
  of purely chance alignments in the galaxy distribution. The analysis
  presented here has distinct advantages over the earlier analysis
  using the LCRS \citep{bharad2}. The  SDSS strips are
  substantially  larger than the  LCRS, and hence there are more
  blocks   which can be shuffled around giving us better statistics
  for the results. Further, the LCRS wedges  were curved and had
  varying thickness whereas our SDSS subsamples are  flat  and have
  uniform thickness.  Our results are consistent with the earlier
  findings for the LCRS where filamentarity was found to be
  statistically significant on scales up to $70-80 \, h^{-1} \, {\rm
  Mpc}$ in the $-3^{\circ}$ slice and  $50-70 \, h^{-1} \, {\rm
  Mpc}$ in the other 5 slices. It is interesting to note that the
  results of the analysis of  mock SDSS catalogues based on \lcdm
N-body simulations \citep{sheth2} reveal
  the length of the longest superclusters to be $  \sim 60 \, h^{-1}
  \, {\rm Mpc}$. A similar analysis of  the supercluster-void
  network in VIRGO \lcdm  N-body simulations show the most massive
  superclusters to exceed $50 \, h^{-1} \, {\rm Mpc}$ in length.
  It may be noted that our analysis gives
  an upper limit to the linear length-scale up to which the filaments
  are statistically significant. The individual filaments may be
  wiggly or coiled up,  and the length measured along  the filament
  may be larger. 

The connectivity and filamentarity of the two SDSS strips are roughly
consistent with each other, though we have not performed a quantitative
comparison given the different geometries of the two subsamples. It
may be noted that the filaments identified in our 2D analysis may
actually be the intersection of 3D planar structures.   Analysis of the
SDSS power-spectrum \citep{teg2} shows the presence of a bump  at the
Fourier mode  $k \sim 0.05 \, h \, {\rm Mpc}^{-1}$  in the power
spectrum.  While it may be speculated that the high level of
filamentarity detected in the SDSS may be a consequence of this
features, earlier analysis using N-body simulations show the
filamentarity to be insensitive to the presence of such a bump
\citep{bharad3}.  

In conclusion we note that our results confirm the earlier results 
\citep{bharad2}  that the filaments seen in the galaxy distribution
are the largest known statistically significant  structures in the
universe.

\section{Acknowledgment}
SB would like to acknowledge financial support from the Govt.
of India, Department of Science and Technology (SP/S2/K-05/2001). BP
would like to thank the CSIR, Govt. of India for financial support
through a JRF fellowship. The authors would like to thank Scot
Kleinman,  Michael Blanton  and  Sebastian Jester for help in 
accessing and understanding the SDSS data. The SDSS DR1 data was
downloaded from the SDSS skyserver http://skyserver.sdss.org/dr1/en/. 

    Funding for the creation and distribution of the SDSS Archive has been 
provided by the Alfred P. Sloan Foundation, the Participating 
Institutions, the National Aeronautics and Space Administration, the 
National Science Foundation, the U.S. Department of Energy, the Japanese 
Monbukagakusho, and the Max Planck Society. The SDSS Web site is 
http://www.sdss.org/.

    The SDSS is managed by the Astrophysical Research Consortium (ARC) for 
the Participating Institutions. The Participating Institutions are The 
University of Chicago, Fermilab, the Institute for Advanced Study, the 
Japan Participation Group, The Johns Hopkins University, the Korean 
Scientist Group, Los Alamos National Laboratory, the Max-Planck-Institute 
for Astronomy (MPIA), the Max-Planck-Institute for Astrophysics (MPA), New 
Mexico State University, University of Pittsburgh, Princeton University, 
the United States Naval Observatory, and the University of Washington.

\end{document}